\documentclass[a4paper,aps,pra,twocolumn,preprintnumbers,amsmath,amssymb]{revtex4}
\usepackage{amsmath}
\usepackage{verbatim}
\usepackage{graphicx}

\usepackage{color}

\begin{document}
 
\title{Excitations and correlations in the driven-dissipative Bose-Hubbard model}

\author{Tobias Gra{\ss}}
\affiliation{Joint Quantum Institute, University of Maryland and NIST, College Park, MD 20742, U.S.A.}

\begin{abstract}
Using a field-theoretic approach within the Schwinger-Keldysh formalism, we study a Bose-Hubbard model in the presence of a driving field and dissipation due to one-body losses. We recover the bistability diagram from the Gross-Pitaevski equation, and analyze the different phases with respect to their elementary excitations and correlations. We find the low-density solution to be subdivided into a dynamically instable, a gapped, and a gapless regime. The correlations decay exponentially, but a substantial increase of correlation length marks the regime of gapless excitations.
\end{abstract}

\maketitle

\section{ Introduction }
The Bose-Hubbard model is a paradigmatic model to describe a quantum phase transition, as site-to-site tunneling processes and on-site interaction processes compete energetically with each other \cite{fisher89}. In a groundbreaking experiment with cold atoms in an optical lattice, the Bose-Hubbard Hamiltonian was realized in a quantum simulator in 2002 \cite{greiner02}. Since then, there has been strong interest in exploiting light-matter interactions to implement also a photonic version of the Bose-Hubbard model \cite{hartmann06,greentree06,angelakis07}. The hopping term is then implemented as an array of coupled cavities, and the presence of either a real atom or an artificial atom in each cavity achieves the non-linearity of the model. A recent realization consists of a linear array of 72 cavities which are coupled to superconducting qubits \cite{fitzpatrick17}. Another realization of a Bose-Hubbard dimer, that is a system of two coupled cavities, has been described in Ref. \cite{rodriguez16}, using polaritons in a semiconductor to obtain the non-linear term. Due to the finite lifetime of the photons or polaritons, these systems do not conserve particle number. To avoid ending up in the vacuum, the losses have to be compensated by an appropriate driving term. A Bose-Hubbard model which is enriched by these two terms is usually refered to as the ``driven-dissipative Bose-Hubbard model'', and has been studied in Refs. \cite{carusotto09,leboite13,leboite14,wilson16,cao16,casteels17,mamaev18,vicentini18,hannukainen18,biondi18}. 

As the driving term explicitly breaks the U(1) symmetry, no quantum phase transition via spontaneous symmetry breaking occurs in the driven model. Nevertheless, the model is not featureless: In particular, it has been shown that the coupling between cavities induces a transition from a monostable to a bistable regime \cite{leboite13,leboite14}, although the variational principle used in Ref. \cite{weimer15} might suggest that the bistability is a mean-field artefact. 
On the other hand, renormalization group arguments for the XX model presented in Ref. \cite{maghrebi16}, as well as an interpretation of the full quantum dynamics in terms of non-equilibrium Langevin equations \cite{foss-feig17}, suggest that the driven-dissipative model falls into the same universality class as the equilibrium Ising model. In this light, the bistability is interpreted as the two minima of the Ising free energy, providing a non-equilibrium analog of the symmetry-broken phase.

In the present paper, we extend the mean-field treatment of the driven-dissipative Bose-Hubbard model, and calculate two-body correlators in order to distinguish between different parameter regimes. As a starting point, we employ a field-theoretic approach on the Keldysh time contour, which is suited to capture the dynamics of the corresponding master equation \cite{sieberer16}. By suppressing all fluctuations, this approach reproduces the results obtained from solving the dissipative Gross-Pitaevski equation, including also the bistability transition. We then admit for fluctuations up to quadratic order, and calculate the Green functions. With this, we gain new insight in the two different steady states. In particular, we calculate the momentum distribution, and we show that the low-density state has a peak structure different from the high-density phase. Studying the correlations in real space, we find an exponential decay in any regime, that is, no true long-range order is established in the system. However, an abrupt increase in the correlation length is seen when the hopping parameter is tuned. The division of the low-density phase into a short-ranged and a longer-ranged regime is also reflected by the elementary excitations of the system, as in the regime of increased correlation length the real part of the excitation spectrum vanishes at certain  momenta.

The paper is organized as follows: Sec. II describes the model, Sec. III introduces the Schwinger-Keldysh method, and Sec. IV presents the results.

\section{ Model }
The Hamiltonian part of the system dynamics consists of the standard Bose-Hubbard model $H_{\rm BH}$ and a driving term $H_{\rm drive}(t)$:
\begin{align}
 H_{\rm BH} = - J \sum_{\langle i,j\rangle} b_i^\dagger b_j + \sum_i \left( \frac{U}{2} \sum_i b_i^\dagger b_i^\dagger b_i b_i + \omega_{\rm c} b_i^\dagger b_i \right),
\end{align}
and
\begin{align}
 H_{\rm drive}(t)= \sum_i \left( j b_i^\dagger {\rm e}^{i \omega_{\rm p} t}+ j^* b_i {\rm e}^{-i \omega_{\rm p} t} \right).
\end{align}
Here, $J$ denotes the hopping strength, $U$ the interaction strength. The cavity frequency $\omega_{\rm c}$ acts like a chemical potential. The pump laser has an amplitude $j$ and a frequency $\omega_{\rm p}$. In a frame rotating with the pump frequency $\omega_{\rm p}$, the Hamiltonin becomes time-independent:
\begin{align}
 H =& - J \sum_{\langle i,j\rangle} b_i^\dagger b_j + \sum_i \Big( \frac{U}{2}  b_i^\dagger b_i^\dagger b_i b_i - \mu b_i^\dagger b_i 
 + j b_i^\dagger + j^* b_i \Big).
\end{align}
Here, the on-site potential term is shifted by the pump frequency, $\mu \equiv \omega_{\rm p} - \omega_{\rm c} $.

Apart from the Hamiltonian dynamics, we will also take into account dissipative particle losses. The full system dynamics is then captured by a master equation
\begin{align}
\label{master}
 i \partial_t \rho = [H,\rho] + \frac{i\gamma}{2} \sum_i \left( 2 b_i\rho b_i^\dagger - b_i^\dagger b_i \rho - \rho b_i^\dagger b_i \right),
\end{align}
with the dissipation rate $\gamma$.

\section{ Schwinger-Keldysh action }

In contrast to the time evolution of state vectors, the quantum-mechanical time evolution of an operator consists of a forward evolution and a backward evolution, $O(t) = U(t,0)^\dagger O(0) U(t,0)$. Therefore, any dynamics which goes beyond the evolution of pure state vectors requires treatment on a closed time contour. Accordingly, the Schwinger-Keldysh action is defined on two time branches, and for any quantity which enters the action we have to specify whether it lies on the forward or on the backward branch. Therefore, let us denote the complex-valued fields on the two branches by $b_{i+}(t)$ and $b_{i-}(t)$. The Schwinger-Keldysh action corresponding to Eq. (\ref{master}) is written as
\begin{widetext}
\begin{align}
 S &= \int{\rm d}t \Big\{ \sum_i \Big[ b_{i+}^*(t) (i\partial_t + \mu) b_{i+}(t) - b_{i-}^*(t) (i\partial_t + \mu) b_{i-}(t) 
- \frac{U}{2} \left( |b_{i+}(t)|^4 - |b_{i-}(t)|^4 \right) - j \left(b_{i+}^*(t) - b_{i-}^*(t) + {\rm c.c.} \right)
\nonumber \\ &
- \frac{i \gamma}{2} \left( 2 b_{i+}(t) b_{i-}^*(t) - b_{i+}^*(t)b_{i+}(t) - b_{i-}^*(t)b_{i-}(t) \right) \Big]
+ J \sum_{\langle i,j \rangle} \left(b_{i+}^*(t) b_{j+}(t) - b_{i-}^*(t) b_{j-}(t) \right) \Big\}.
 \end{align}

It is convenient to rotate the fields to so-called classical fields $b_{i,{\rm c}}(t) = \frac{1}{\sqrt{2}}\left[ b_{i+}(t) + b_{i-}(t) \right]$, and quantum fields $b_{i,{\rm q}}(t) = \frac{1}{\sqrt{2}}\left[ b_{i+}(t) - b_{i-}(t) \right]$. In this basis, after integrating by parts a term containing $\partial_t b_{i,{\rm q}}$, the action reads
\begin{align}
 S &= \int{\rm d}t \Big\{ \sum_i \Big[ b_{i,{\rm q}}^* \left(i \partial_t +\mu -\frac{i\gamma}{2} \right) b_{i,{\rm c}} - \sqrt{2}j b_{i,{\rm q}}^* 
 - \frac{U}{2} b_{i{\rm q}}^* b_{i,{\rm c}} \left( |b_{i,{\rm c}}|^2 + |b_{i,{\rm q}}|^2 \right) + \frac{i \gamma}{2}|b_{i,{\rm q}}|^2 \Big] + {\rm c.c.}
 \nonumber \\ &
+ J \sum_{\langle i,j \rangle} \left( b_{i,{\rm c}}^* b_{j,{\rm q}} + b_{i,{\rm q}}^* b_{j,{\rm c}} \right) \Big\}.
\end{align}
For brevity, we have suppressed the time-dependence of the fields.
\end{widetext}

\subsection{Mean-field solution}

A mean-field solution to the equations of motion,
\begin{align}
 \frac{\delta S}{\delta b_{i,{\rm c}}^*} = 0 \ \ \ \ {\rm and} \ \ \ \ \frac{\delta S}{\delta b_{i,{\rm q}}^*} = 0,
\end{align}
is found by setting 
\begin{align}
 b_{i,{\rm q}}=0, \ \ \ \ {\rm and} \ \ \ \
 b_{i,{\rm c}}=\sqrt{2} b,
\end{align}
where $b$ has to fulfill a Gross-Pitaevski-like equation:
\begin{align}
\label{gp}
 b \left[ \mu - U |b|^2 + zJ - \frac{i\gamma}{2} \right] - j = 0.
\end{align}
The same equation has been formulated in Ref. \cite{leboite14}. Multiplied by its complex conjugate, it leads to a bistability criterion:
\begin{align}
\label{n}
 n \left[ ( \mu - n U + zJ)^2 +\frac{\gamma^2}{4} \right] = |j|^2,
\end{align}
where the mean photon number $n=|b|^2$ needs to be real. From this equation, a monostable phase is predicted when only one real solution exists, which is the case for small $J/\mu$. For larger $J/\mu$, Eq. (\ref{n}) admits admits three real solutions. A comparison with the $P$-representation approach suggests that only two of them are stable \cite{leboite14}.  Such comparison also suggests that the Gross-Pitaevski approximation underestimates the extension of the monostable regime \cite{leboite14}.

\subsection{Fluctuations around mean-field}

In the previous section, all fluctuations have been neglected. In the next step, we will take them into account up to second order. We write
\begin{align}
 b_{i,{\rm q}}(t)=\delta_{i,{\rm q}}(t), \ \ \ \ {\rm and} \ \ \ \
 b_{i,{\rm c}}(t)=\sqrt{2}b + \delta_{i,{\rm c}}(t).
\end{align}
Via the choice of $b$ from Eq. (\ref{gp}), the first-order contribution to the action vanishes, $S^{(1)}=0$. Thus, to leading order we have an action which is quadratic in $\delta$:
\begin{align}
\label{S2}
 S^{(2)}=& \int {\rm d}t \Bigg\{ \sum_i \Big[ \delta_{i,{\rm q}}^* \left( \partial_t + \mu -\frac{i\gamma}{2} - 2U |b|^2\right) \delta_{i,{\rm c}} 
  \nonumber \\ &
 - U b^2 \delta_{i,{\rm q}}^* \delta_{i,{\rm c}}^*  + \frac{i \gamma}{2} |\delta_{i,{\rm q}}|^2\Big] +
 \nonumber \\ & 
  + J \sum_{\langle i,j\rangle} \left( \delta_{i,{\rm c}}^* \delta_{j,{\rm q}} + \delta_{i,{\rm q}}^* \delta_{j,{\rm c}} \right) 
  \Bigg\}
 + {\rm c.c.} 
\end{align}
In general, the quadratic Schwinger-Keldysh action is given in terms of three Green functions. In Fourier space and after introducing Nambu spinors $\boldsymbol\delta_{{\rm c},{\bf k}}(\omega) = ( \delta_{{\rm c},{\bf k}}(\omega),  \delta_{{\rm c},{-\bf k}}(\omega)^* )^T$, this relation reads
\begin{align}
\label{S2general}
 S^{(2)} = & \int {\rm d}\omega \sum_{\rm k} (\boldsymbol\delta_{{\rm c},{\bf k}}(\omega)^*, \boldsymbol\delta_{{\rm q},{\bf k}}(\omega)^* ) \cdot
\nonumber \\ &
 \left( 
 \begin{matrix}
  0 & P^{\rm A}(\omega,{\bf k}) \\
  P^{\rm R}(\omega,{\bf k}) & P^{\rm K}(\omega,{\bf k})
 \end{matrix}
 \right)
 \cdot
 \left(
\begin{matrix}
 \boldsymbol\delta_{{\rm c},{\bf k}}(\omega) \\
 \boldsymbol\delta_{{\rm q},{\bf k}}(\omega)
\end{matrix}
\right)
\end{align}
Here, $P^A(\omega,{\bf k})=[G^A(\omega,{\bf k})]^{-1}$ and $P^R(\omega,{\bf k})=[G^R(\omega,{\bf k})]^{-1}$ are the inverse of the advanced and the retarded Green functions:
\begin{align}
& G^R(\omega,{\bf k})= G^A(\omega,{\bf k})^\dagger =  \nonumber \\ &
 -i \left(
\begin{matrix}
 \langle \delta_{{\rm c},{\bf k}}(\omega) \delta_{{\rm q},{\bf k}}(\omega)^*  \rangle &  \langle \delta_{{\rm c},{\bf k}}(\omega) \delta_{{\rm q},{-\bf k}}(\omega)  \rangle \\
 \langle \delta_{{\rm c},{-\bf k}}(\omega)^* \delta_{{\rm q},{\bf k}}(\omega)^*  \rangle &  \langle \delta_{{\rm c},{-\bf k}}(\omega)^* \delta_{{\rm q},{-\bf k}}(\omega)  \rangle 
\end{matrix}
\right).
\end{align}
From the retarded or advanced Green function, we obtain the excitation spectrum by solving for ${\rm det} [G^R(\omega,{\bf k})]^{-1} = 0$.
The function $P^K(\omega,{\bf k})$ is related to the Keldysh Green function $G^K$ in the following way:
\begin{align}
G^K(\omega,{\bf k})=-G^A(\omega,{\bf k}) P^K(\omega,{\bf k}) G^R(\omega,{\bf k}).
\end{align}
The Keldysh Green function contains information about correlation of the classical fields:
\begin{align}
& G^K(\omega,{\bf k})=\nonumber \\ &
-i \left(
\begin{matrix}
 \langle \delta_{{\rm c},{\bf k}}(\omega) \delta_{{\rm c},{\bf k}}(\omega)^*  \rangle &  \langle \delta_{{\rm c},{\bf k}}(\omega) \delta_{{\rm c},{-\bf k}}(\omega)  \rangle \\
 \langle \delta_{{\rm c},{-\bf k}}(\omega)^* \delta_{{\rm c},{\bf k}}(\omega)^*  \rangle &  \langle \delta_{{\rm c},{-\bf k}}(\omega)^* \delta_{{\rm c},{-\bf k}}(\omega)  \rangle 
\end{matrix}
\right).
\end{align}
Comparison of the general expression (\ref{S2general}) with the Fourier-transformed action of Eq. (\ref{S2}) yields:
\begin{align}
 P^R(\omega,{\bf k})=
-i \left(
\begin{matrix}
 \omega + C_{\bf k} & -U b^2  \\
 -U b^{*2} &  -\omega + C_{\bf k}^*
\end{matrix}
\right),
\end{align}
where $C_{\bf k} =  \mu +\frac{i\gamma}{2} - 2U|b|^2  + 2 J [\cos(k_x) +\cos(k_y) ]$. The Keldysh component reads
\begin{align}
 P^K(\omega,{\bf k})=
i\gamma \left(
\begin{matrix}
 1 & 0 \\ 0 & 1
\end{matrix}
\right).
\end{align}
The excitation spectrum obtained from the retarded Green function matches with the spectrum found in Ref. \cite{leboite14}:
\begin{align}
\label{ex}
 \omega_\pm({\bf k}) = \pm \sqrt{(\mu+J_{\bf k}-2U|b|^2)^2-U^2|b|^4}-\frac{i\gamma}{2}.
\end{align}
From the diagonal elements of the Keldysh Green function, we obtain the momentum distribution:
\begin{align}
n_{\bf k} \equiv \langle \delta_{{\bf k}} \delta_{{\bf k}}^* \rangle  = \frac{i}{4\pi} \int G^K_{11}(\omega,{\bf k}) {\rm d}\omega -\frac{1}{2}.
\end{align}
The off-diagonal elements are related to superfluid correlations:
\begin{align}
\Delta_{\bf k} \equiv \langle \delta_{{\bf k}} \delta_{{\bf k}} \rangle = \frac{i}{4\pi} \int G^K_{12}(\omega,{\bf k}) {\rm d}\omega,
 \\
\Delta_{\bf k}^* = \langle \delta_{{\bf k}}^* \delta_{{\bf k}}^* \rangle = \frac{i}{4\pi} \int G^K_{21}(\omega,{\bf k}) {\rm d}\omega .
\end{align}
Note that both $n_{\bf k}$ and $\Delta_{\bf k}$ are defined as correlators between the \textit{fluctuating} part $\delta$ of the fields. At ${\bf k}={\bf 0}$, we have a contribution from the mean-field part:
\begin{align}
\label{tilden}
 \tilde n_{\bf 0} \equiv \langle (b+\delta_{{\bf 0}})(b^*+ \delta_{{\bf 0}}^*) \rangle = |b|^2+n_{\bf 0},
\end{align}
and
\begin{align}
\label{tildedelta}
  \tilde \Delta_{\bf 0} \equiv \langle (b+\delta_{{\bf 0}})(b+ \delta_{{\bf 0}}) \rangle = |b|^2+\Delta_{\bf 0},
\end{align}
as $\langle \delta_{{\bf 0}} \rangle =0$.

\section{Results} 

\begin{figure}
\centering
\includegraphics[width=0.48\textwidth, angle=0]{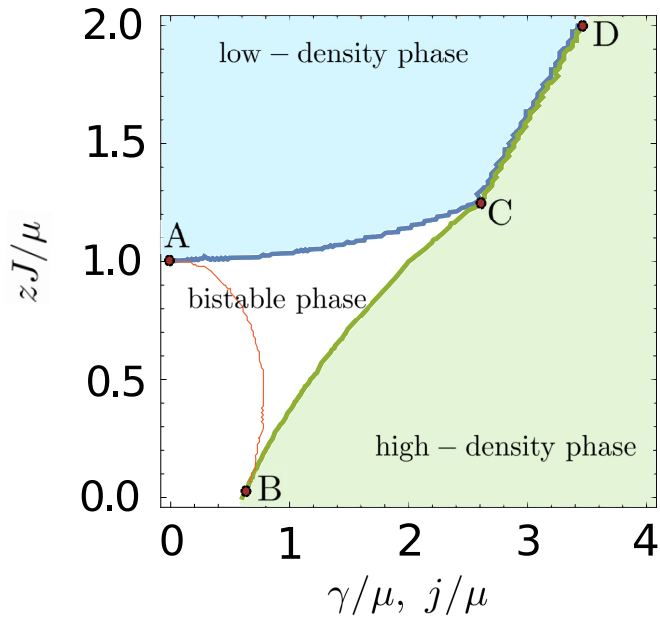}
\caption{\label{fig_bistab} 
We plot the phase diagram of the driven-dissipative Bose-Hubbard model, obtained from a homogeneous mean-field ansatz. We have fixed the ratio $U/\mu=0.5$, which leaves three adjustable parameters $zJ/\mu$, $\gamma/\mu$, and $j/\mu$. For a two-dimensional plot, we further choose $\gamma=j$. Eq. (\ref{n}) yields three different phases: a low-density phase (blue), a high-density phase (green), and a bistable phase (white). All phases meet in the critical point $C$. Along a critical line from $C$ to $D$, the high-density phase and the low-density phase are identical. The phase boundaries can be obtained by searching for gapless imaginary parts in the excitation spectra, i.e. by demanding ${\rm Im}[\omega(\bf k)]=0$. Along the blue line from $A$ to $C$ and further to $D$, the imaginary part of excitations above the high-density solution becomes gapless. Along the green line from $B$ to $C$ and further to $D$, the imaginary part of excitations above the low-density solution becomes gapless. Within the bistable phase, the real part of excitations above the high-density solution becomes gapless along the red line from A to B.
}
\end{figure}

In the following section we will explicitly evaluate the action derived above, and extract relevant physical quantities. To ease our discussion, we will reduce the parameter space: Of the five energy scales ($\mu$, $U$, $zJ$, $\gamma$, and $j$), one scale can be taken as a unit of energy (here $\mu$), leaving four independent parameters. Since the truncation of the action to quadratic order can only be justified if the effect of interactions is small, we keep $U$ fixed at a small level, $U/\mu=0.5$. In Fig. \ref{fig_bistab}, we have further set $\gamma$ equal to $j$, such that we can plot the two-dimensional parameter space. This plot shows the division into bistable and monostable phases according to Eq. (\ref{n}), where the two solutions within the bistable regime can be distinguished by their density. Interestingly, the phase boundary of the bistable regime can also be obtained from the imaginary part of the excitation spectra, ${\rm Im}[\omega(\bf k)]$, which becomes gapless at the bistability transition. Along the lower branch of the boundary (from $B$ to $C$ in Fig. \ref{fig_bistab}), the vanishing imaginary gap occurs to excitations above the low-density solution. Accordingly, the low-density solutions become instable here, and the system enters a monostable, high-density phase. The opposite is the case along the upper branch of the phase boundary (from $A$ to $C$), with a vanishing gap occuring in the imaginary part of the excitations above the high-density phase, and the system entering a monostable low-density phase. 

The criterion of a vanishing gap in ${\rm Im}[\omega(\bf k)]$ also serves to establish a direct boundary between the two monostable phases. On such critical line (from $C$ to $D$ in Fig. \ref{fig_bistab}), both types of excitations, high- and low-density excitations show not only gapless imaginary but also gapless real parts. Gapless real parts (without gapless imaginary parts) are also found within the bistable phase (red line).

\subsection{Low-density phase} 

\begin{figure}
\centering
\includegraphics[width=0.48\textwidth, angle=0]{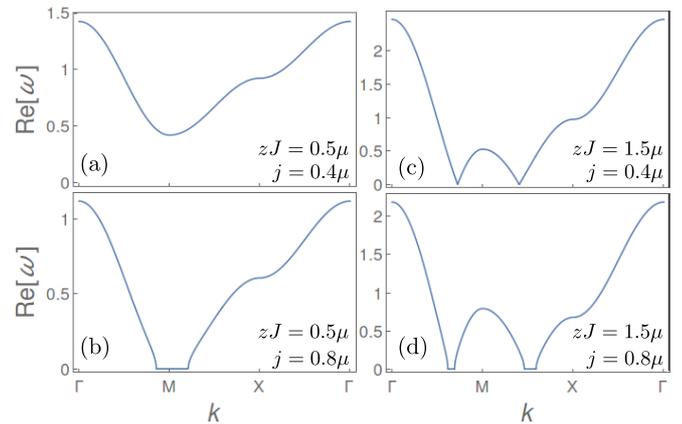}
\caption{\label{fig_ld_spec} Elementary excitations above the low-density steady state are plotted along high symmetry points in the first Brillouin zone. In all panels, the interaction strength is fixed to $U/\mu=0.5$, and the dissipation rate is fixed to $\gamma/\mu=0.2$. The cavity coupling is $zJ\mu=0.5$ in panels (a,b), and $zJ/\mu=1.5$ in panels (c,d). The driving amplitude is $j\mu=0.4$ in panels (a,c), and $j\mu=0.8$ in panels (b,d).
}
\end{figure}

 \begin{figure}[t]
\centering
\includegraphics[width=0.48\textwidth, angle=0]{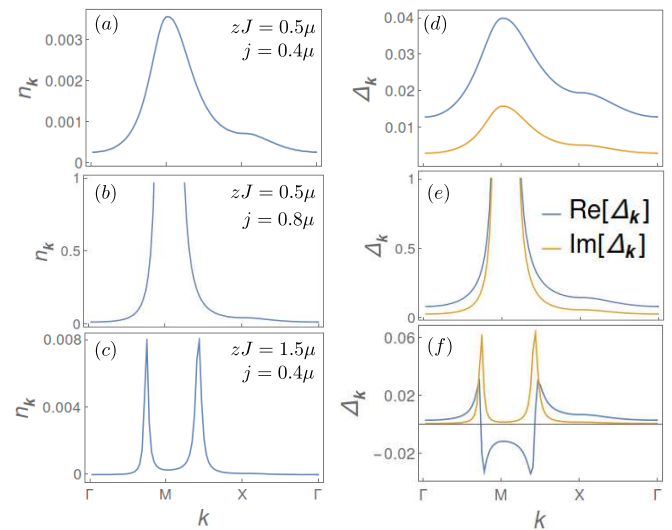}
\caption{\label{fig_ld_nk} Momentum distribution $n_{\bf k}$ (a--c) and superfluid correlations $\Delta_{\bf k}$ (d--f) in the low-density phase for different values of $zJ$ and $j$. In all panels $U\/mu=0.5$, and $\gamma/\mu=0.2$. In (a,d), we have chosen $zJ/\mu=0.5$ and $j/\mu=0.4$, such that the system is in the gapped phase. In (b,e), we have chosen $zJ/\mu=0.5$ and $j/\mu=0.8$, giving an example for a system in the dynamically instable phase. In (c,f), we have chosen $zJ/\mu=1.5$ and $j/\mu=0.4$, bringing the system in the gapless phase. The superfluid correlations exhibit a $\pi$ phase shift around the gapless points.
}
\end{figure}

We first study the elementary excitations obtained by plugging the low-density solution of Eq. (\ref{gp}) into Eq. (\ref{ex}). As illustrated in Fig. \ref{fig_ld_spec}, the spectra vary significantly depending on the choice of parameters $J/\mu$ and $j/\mu$: If both cavity coupling $J$ and driving $j$ are weak, as illustrated in panel (a), the excitations are gapped. The gap can be removed by increasing either $J$ or $j$. As shown in panel (b) and (d), increasing $j$ leads to an extended region in $\bf k$-space in which the real part of the excitation frequency, ${\rm Re}(\omega)$, vanishes. In these regions, the imaginary part of the frequencies, gains a contribution additional to the otherwise constant value $-i\gamma/2$. This constant contribution directly stems from the dissipative nature of the system, and is present for all ${\bf k}$ in any parameter regime. We interpret this increase of the imaginary part as a dynamical instability which occurs even before the low-density solution becomes statically instable, e.g., before the density from Eq. (\ref{n}) becomes imaginary. A stable gapless phase can be reached by increasing the coupling $J$ while keeping $j$ sufficiently weak, as illustrated in panel (c) of Fig. \ref{fig_ld_spec}. In this case, the real part of the excitation frequencies vanishes on a circle around the M-point [${\bf k}a=(\pi,\pi)$], with linear dispersion relation around these gapless points. In summary, our analysis allows for distinguishing between four regions in parameter space: region of statical instability, region of dynamic instability, and a stable region divided into a gapped and a gapless regime (with respect to the real part of the excitation frequency).

 \begin{figure}
\centering
\includegraphics[width=0.48\textwidth, angle=0]{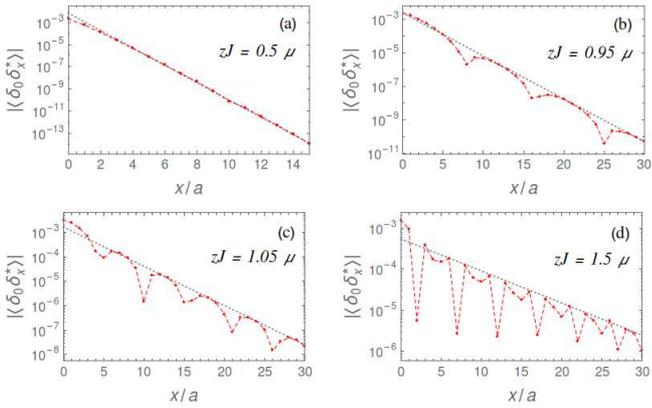}
\caption{\label{fig_spacecorr} For three different values of $zJ/\mu$ the spatial behavior of correlations $|\langle \delta_0 \delta_x\rangle|$ along the lattice axis is plotted:
(a) For $zJ=0.5\mu$, correlations decay exponentially with a short correlation length $\xi/a \lesssim 1$. For larger values of $zJ$, the exponential decay becomes spatially modulated, reflecting a condensation of fluctuations at non-zero momentum. While in (b), for $zJ=0.95\mu$, this modulation is still weak, it becomes more and more pronounced in (c) for $zJ=1.05\mu$, and (d) for $zJ=1.5\mu$. From (a) to (d), the correlation length is increased by one order of magnitude. In all panels, we have chosen $U=0.5\mu$ and $\gamma=0.2\mu$. The driving amplitude is set to $j=0.2\mu$.}
\end{figure}

 \begin{figure}
\centering
\includegraphics[width=0.40\textwidth, angle=0]{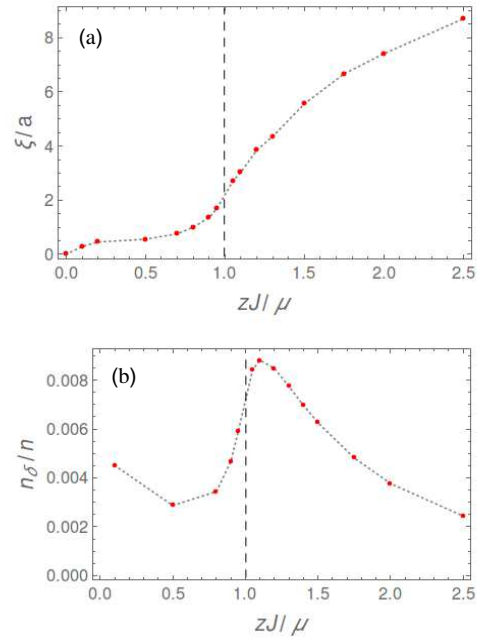}
\caption{\label{fig_xidep} We plot (a) the correlation length $\xi$ and (b) the depletion ratio $n_\delta/n$ as a function of $zJ/\mu$ for fixed values of  $j=0.2\mu$, $U=0.5\mu$, and $\gamma=0.2\mu$. While for small $zJ<\mu$ (e.g. left to the dashed vertical line), the correlation length is of the order 1, it increases significantly for $zJ>\mu$. The depletion ratio in (b) is small for all values $zJ$, justifying the truncation of the action to quadratic terms. The dynamical instability around $zJ \sim 1$ leads to a peak in the depletion in that regime.}
\end{figure}

From the Keldysh component of the Green function, we also obtain information about the momentum distribution and superfluid correlations. Not surprisingly, they directly reflect the different properties of the excitation spectra. As seen in Fig. \ref{fig_ld_nk}, maxima in $n_{\bf k}$ and $|\Delta_{\bf k}|$ correspond to those points in ${\bf k}$-space where the real part of the excitation spectra exhibits minima, cf. Fig. \ref{fig_ld_spec}. In the gapless phase, these maxima become sharply peaked. Divergencies in $n_{\bf k}$ and $|\Delta_{\bf k}|$ characterize the dynamically instable phase. Interestingly, upon tuning through a gapless point, the superfluid correlations $\Delta_{\bf k}$ exhibit a $\pi$-phase shift, see Fig. \ref{fig_ld_nk}(f).

Transforming back to spatio-temporal variables, we further analyze the range of the equal-time correlations:
\begin{align}
 \langle \delta_{\bf r} \delta_{{\bf r}+{\bf r'}}^* \rangle = \frac{1}{(2\pi)^2} \sum_{\bf k} e^{i{\bf k}\cdot({\bf r}-{\bf r'})} n_{\bf k}.
\end{align}
The range of correlations is then quantified by the correlation length $\xi$ which we obtain by fitting the decay of correlations along a lattice direction to an exponential function:
\begin{align}
|\langle \delta_0 \delta_{x}^* \rangle| \propto \exp(-x/\xi).
\end{align}
We have performed this analysis at $\gamma=0.2\mu$ and $U=0.5\mu$ in a weakly driven regime, $j=0.2\mu$, scanning several values of the cavity coupling $zJ/\mu$ in the interval $(0,2.5]$. According to our previous results, we expect quantitative and qualitative changes of the correlations near $zJ=\mu$. Indeed, the behavior can be described in the following way: For $zJ<\mu$, the spatial decay of is very well captured by the exponential fit, as seen in Fig. \ref{fig_spacecorr}(a). When approaching $zJ\approx 1$ and for even larger values of $zJ$, the exponential decay is superposed by a spatial modulation of correlation, see Fig. \ref{fig_spacecorr}(b--d). At the same time, the correlation length $\xi$ starts to increase, with ${\rm d}\xi/{\rm d}J$ being largest at $zJ/\mu=1$. This behavior is illustrated in Fig. \ref{fig_xidep}(a). This observation supports the notion of two distinct regimes: At weak cavity coupling, $zJ<\mu$, the system exhibits only short-ranged correlations, while for $zJ>\mu$ the correlations acquire significantly longer range.

Finally, we have a look on the depletion. In order to justify the truncation of the action to second-order in $\delta$, we have to demand that the ratio between the density of flucuating particles, $n_\delta = \langle \delta_0 \delta_0\rangle$, and the density of condensed particles $n=|b|^2$ is small. As seen in Fig. \ref{fig_xidep}(b), at low drive $j=0.2\mu$, we have $n_\delta /n  \lesssim 0.01$ for all values of $zJ/\mu$, rendering our description self-consistent. We also note that $n_\delta /n$ has a maximum near the critical coupling strength $zJ/\mu=1$ which might reflect a dynamical instability.

\begin{figure*}
\centering
\includegraphics[width=0.98\textwidth, angle=0]{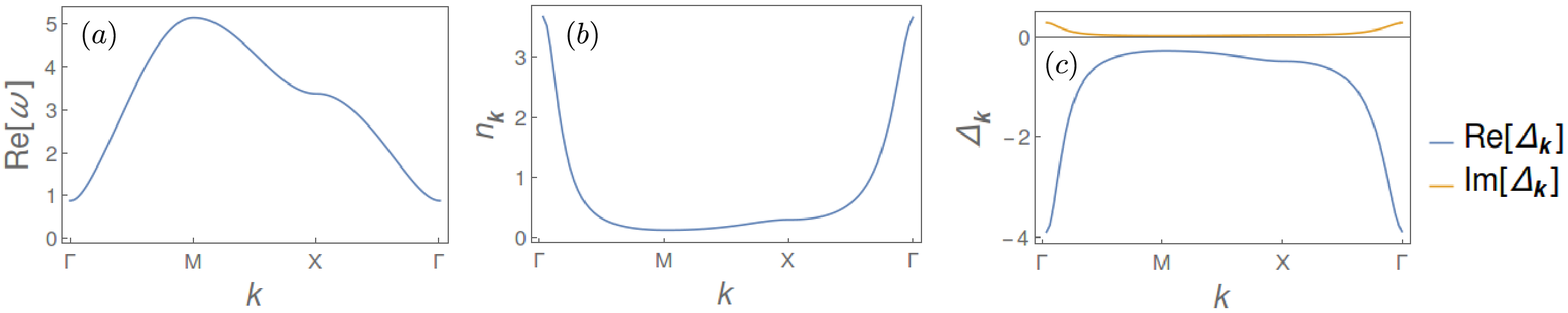}[h]
\caption{\label{fig_hd} Elementary excitations (a), momentum distribution $n_{\bf k}$ (b), and superfluid correlations $\Delta_{\bf k}$ (c) in the high-density phase, exemplified for $zJ=U=0.5\mu$, $\gamma=0.2\mu$, and $j=0.4\mu$.}
\end{figure*}

\subsection{High-density phase} 
In contrast to the low-density phase, the high-density solution to Eq. (\ref{n}) does not appear to be further divided into different subphases. For any choice of parameters, we find gapped elementary excitations as illustrated in Fig. \ref{fig_hd} (a). Again, the imaginary part of the excitation frequencies is always pinned to $i\gamma/2$. In contrast to the gapped phase at low density, with its lowest excitation at the M-point in the corner of the first Brillouin zone, the high density phase has its lowest excitation at the $\Gamma$-point in the center of the first Brillouin zone. This is also reflected by the position of the maximum of the momentum distribution $n_{\bf k}$ or superfluid correlations $\Delta_{\bf k}$, see Fig. \ref{fig_hd} (b,c).

\section{Conclusions}
We have studied a two-dimensional Bose-Hubbard model with dissipative losses and coherent driving. From the Schwinger-Keldysh action of the system, we obtain equations of motion which have one or several homogeneous solutions, leading to the distinction between mono- and multistable regimes. Fluctuations around these solutions yield the excitation spectra which allow for a further division of the low-density solution into a gapped, a gapless, and a dynamically instable regime. 
While our analysis was based on a mean-field approximation, the assumption of weak fluctuations was self-consistently confirmed. 
Our calculations may be relevant to future experiments with two-dimensional Bose-Hubbard models in a driven-dissipative environment. As an experimentally accessible figure of merit, we have computed the two-body correlations in the system. It is found that the transition from the gapped to the gapless regime is characterized through a significant increase of the correlation length.

\begin{acknowledgments}
 I acknowledge fruitful discussions with Bin Cao, Michael Foss-Feig, and Mohammad Maghrebi. Financial support from the NSF through PFC at JQI is acknowledged.
\end{acknowledgments}


\end{document}